\documentclass[referee]{aa}
\usepackage{graphicx}
\usepackage{txfonts}
\usepackage{natbib}
\newcommand\eg{{e.g.,~}}

\begin{document}
\title{Stellar halos and thick disks around edge-on spiral galaxies
       IC~2233, IC~5052, NGC~4631 and NGC~5023.
\thanks{Based on observations with the NASA/ESA Hubble Space Telescope,
obtained at the Space Telescope Science Institute, which is operated by the
Association of Universities for Research in Astronomy, Inc., under NASA
contract NAS 5-26555.}}
\titlerunning{Stellar outskirts around edge-on galaxies}
\authorrunning{Tikhonov et al.}
\author{Tikhonov N.A. \inst{1,2}, Galazutdinova O.A. \inst{1,2}, Drozdovsky I.O.\inst{3,4}}
\institute{Special Astrophysical Observatory, Russian Academy
   of Sciences, N.Arkhyz, KChR, 369167, Russia
\and Isaac Newton Institute of Chile, SAO Branch, Russia
\and Instituto de Astrof\'{\i}sica de Canarias, E--38200, Tenerife, Spain
\and Astronomical Institute, St.Petersburg University, 198504, Russia}
\date {Received January 30 , 2006; \ accepted  }
\abstract
{}
{The extraplanar stellar populations of highly-inclined
disk galaxies IC~2233, IC~5052, NGC~4631 and NGC~5023
are analyzed with the goal to quantify their vertical extent
and structure.}
{Based on the single-star photometry, we separate different stellar
populations
and analyze their spatial distribution.}
{On archival images obtained with the Hubble Space Telescope ACS/WFC
the surroundings of these galaxies are well resolved into stars with
the red giant population (RGB) identified far above the galaxy mid-planes.
We find that there are a profound change in slope of the number density
profile of the evolved RGB stars at extraplanar height of $4-8$~kpc,
that possibly reflects a truncation of the thick disk and reaching the
$2-3$ times more extended oblate stellar component (``the halo'').
This structure is consistent with our previous studies of both
edge-on and face-on disk galaxies in the Local Universe (Tikhonov et al. 2005a,b)
and allow us to improve the spatial model of the stellar components of
a typical spiral galaxy.
In NGC~4631, the revealed asymmetry of its stellar thick disk and halo,
is likely caused by the neighbor dwarf galaxy NGC~4627.
Based on the tip of the red giant branch
method (TRGB) we estimated a distance of $10.42\pm0.38$~Mpc for IC~2233,
$5.62\pm0.20$~Mpc for IC~5052, $7.11\pm0.13$~Mpc for NGC~4631 ($6.70\pm0.15$
for its satellite NGC~4627) and $6.14\pm0.15$~Mpc for NGC~5023.
We confirm the presence of slight extraplanar metallicity
gradient of evolved stars at NGC~5023 and IC~5052, based on the systematic
changes of the colour distribution of red giant stars.}
{}

\maketitle

\section{Introduction}

Understanding the structure of the stellar component of disk galaxies
is an important step for unraveling their formation and evolution.
Although challenging to measure, the extraplanar stellar content
around galaxies is a potentially-important probe of the most ancient
stellar populations, as well as of the role of that different
external and internal processes play in the evolution of modern galaxies,
such as: merging, accretion and infall of metal-poor gas and satellite
galaxies \citep[\eg][]{Abadi:03,Brook:04,Bullock:05}; disk heating
\citep[\eg][]{Kroupa:02,Gnedin:03}; or the ejection of
of gas from the underlying disk, which may cool so that stars can form
\citep[\eg][]{Howk:99,Savage:03}.

The stellar outskirts around our own Galaxy and some
other large spiral galaxies has been extensively studied
over the last few years and revealed faint stellar envelopes in
many of them \citep[\eg][]{Sackett:99, Cuillandre:99, Marel:01,
Ferguson:02, Dalcanton:02, Zibetti:04b}.
The most extensive and detailed study of the outer stellar
populations of the Milky Way and M~31 revealed complex patchy structures with
numerous streams, which appear to be the remnants of disrupted
companion galaxies \citep[\eg][]{Brown:03, Zucker:04}.
However, the chemical patterns of the stars of the dwarf spheroidal (dSph),
and irregular (LMC and SMC) satellite galaxies are
predominantly distinct
from the stars in each of the kinematic components of the Galaxy.
This result rules out continuous merging
of low-mass galaxies similar to the currently observed satellites
during the formation of the Milky Way
\citep[\eg][]{SHane:98,McWilliam:05,Venn:04}.
At the same time, the detection of extraplanar cold and ionized gas regions
at large distances from the mid-planes of some spiral
galaxies, which must be powered by hot stars, suggests that
while most of the outer, elusive stellar components are predominantly
evolved, some fraction of these stars are not
a single burst ancient stellar population
\citep[\eg][]{Keppel:91,Hoopes:99,Howk:99,Tullmann:03}.

In the literature, the definition ``halo'' and ``thick disk'' is rather
unclear, as well as the method of their separation are not well established
yet.
While some of the recent studies of the outer regions of spiral galaxies
rely on multicolour surface photometry, successful detection is only possible
in the deep observations of the edge-on galaxies with a prominent thick disk
and halo \citep{Dalcanton:02, Pohlen:04, Zibetti:04a, Zibetti:04b}.
For example, \citet{Dalcanton:02} analyzed the sample of 47
edge-on spiral galaxies and have found a presence
of the thick disk in  90\% of them, based on the results of multicolour
surface photometry. By stacking of 1054 rescaled images of
edge-on galaxies in the Sloan DSS \citet{Zibetti:04a} detected
an extended low surface brightness
halo, however the ``mean'' colours of this component
($g-r=0.65$ and $r-i=0.6$)
cannot be easily explained by normal stellar populations, and may suggest
contamination by ionized gas emission.
The important other result obtained at surface photometry studies
is that both thin and thick disks are truncated, with mean scaleheight
and scalelength of the thick disk is several times larger than that of
the thin disk (Pohlen et al. 2004).
Given that there have also been some non-detections of luminous thick disk
or halo components \citep[e.g][]{Fry:99,Zheng:99}, the
question arises about the ubiquity, origin and further evolution of these
outer structures.

The alternative technique of
tracking out resolved stars to explore various
galactic subsystems sidesteps the difficulties inherent in quantifying
extremely faint diffuse emission.
The measurement of single-star metallicity and velocity fields
can only be applied to distinguish
different stellar components in Milky Way and in a handful of nearest
galaxies \citep{Gilmore:83, Chiba:00, Prochaska:00, Worthey:05, Ibata:05}. 
For the more
distant galaxies this information is unavailable, and a
set of the model dependent assumptions need to be considered.
Aside from the spatial distribution law and kinematics,
stars from each of the galactic subsystems (bulge, disks, and halo) share
a common star formation history
retaining considerable age and chemical information
\citep{Vallenary:00, Prochaska:00, Chiba:00, Sarajedini:01, Williams:02,
Brewer:04}. On the basis of multi-colour single-star photometry, one is
able to separate different stellar populations, and suggesting
that each of the components has a different
dominant stellar population, to constrain their spatial
distribution.

Inspired by the unparalleled angular resolution and sensitivity
of the {\it HST} WFPC2 and ACS, we began a project to
analyze the outermost stellar structure of spiral galaxies.
In related works,
we have analyzed the the spatial distribution and properties of
the outer stars
in the nearby face-on (\object{M~81} and \object{NGC~300}), as well as
edge-on
(\object{NGC~55}, \object{NGC~891}, \object{NGC~4144}, and \object{NGC~4244})
spiral galaxies \citep{ntik:05a, ntik:05b}.
In the most massive of the studied disk galaxies
we have found a profound change in slope of the
number density profile of the evolved red giant branch (RGB) stars,
that likely reflects a truncation of the
the thick disk and reaching the pure halo
stellar component.
The similar change of the number density gradient of RGB stars were also
reported for lower mass spiral M~33 \citep{Cuillandre:99, Rowe:05}, as
well as in two massive irregular galaxies: 
\object{IC~10} \citep{dio:03}, and M~82 \citep{ntik:05c}.
These results strongly support the ubiquitous presence of the truncated
thick disks in the majority of disk galaxies, as well as large
extend of the stellar outskirts in all directions.

In this paper, we
selected four high-inclination nearby disk galaxies
for an in-depth study
of their extraplanar stellar structure: IC~2233, IC~5052, NGC~4631 and
NGC~5023. Our primary objectives are to: (i) quantify the stellar population
variations associated with extreme outer substructure, and
(ii) derive stringent constraints on the age and metallicity of stars
in the far outer disk.
The choice of high-inclination galaxies ensures minimal contamination
by stars belonging to the dominant galactic disk component, and
provides the more robust separation of the thick disk and  halo
stellar populations.

\section{Observations and stellar photometry}

\begin{figure*}[tb]
\vbox{\includegraphics{fig1_4.ps}
}

\vspace{16.5cm}
\caption{The DSS-2 images of the galaxies with the ACS/WFC footprints
superposed. The edge of thick disk and halo extent are shown by the ellipses.
}
\label{f:Ima}
\end{figure*}

\begin{table}
\begin{center}
\footnotesize
\caption{The ACS/WFC Survey Fields}\label{t:obslog}
\renewcommand{\tabcolsep}{2pt}
\begin{tabular}{lcccccr}\\ \hline \hline
\multicolumn{1}{c}{Galaxy}&
\multicolumn{1}{c}{Region}&
\multicolumn{1}{c}{Date}&
\multicolumn{1}{c}{Band}&
\multicolumn{1}{c}{Exposure}&
\multicolumn{1}{c}{ID}&
\multicolumn{1}{r}{N$_{stars}$ }\\
			    \hline\\
IC2233 & S1   & 2004-04-29 & F814w& 2$\times$350  & 9765 &27120\\
       &      & 2004-04-29 & F606w& 2$\times$338  & 9765 &\\
NGC4631& S1   & 2003-08-03 & F814w& 2$\times$350  & 9765 &115315\\
       &      & 2003-08-03 & F606w& 2$\times$338  & 9765 &\\
       &  S2  & 2004-06-09 & F814w&  2$\times$350  & 9765 &108980\\
       &      & 2004-06-09 & F606w&  2$\times$338  & 9765 &\\
IC5052 &  S1  & 2003-12-13 & F814w&  2$\times$350  & 9765 &63108\\
       &      & 2003-12-13 & F606w&  2$\times$338  & 9765 &\\
NGC5023&  S1  & 2004-07-02 & F814w&  2$\times$350  & 9765 &49628\\
       &      & 2004-07-02 & F606w&  2$\times$338 & 9765 &\\

\hline
\end{tabular}
\end{center}
\end{table}

To study the resolved stellar population of galaxies, we obtained
ACS/WFC images available in the {\it HST\/} archive.
Digital Sky Survey images of these galaxies with ACS/WFC footprints are
shown in Fig.~\ref{f:Ima}.
An interior and exterior ellipses delineate an estimated edges of the thick
disk and halo.
The observational data are listed in Table~\ref{t:obslog},
where ID is the HST program number and N$_{stars}$ is the
number of detected stars.
Images were reprocessed through the standard
ACS STScI pipeline, {\tt CALACS}, as described by \citet{Pavlovsky:05}.
All data have first
order bias subtraction, dark subtraction, and bad pixel masking applied.
Images are then corrected for the flat-field response.
We combined images using a latest version of the {\tt
Multidrizzle} package \citep{Koekemoer:02}, which provides
an automated method for distortion-correcting and combining dithered
images. {\tt Multidrizzle} also corrects for gain and bias offsets between
WFC chips, identifies and removes cosmic rays and cosmetic defects.
We generate point-source catalogs from the stacked direct image using an
automatic star-finding program  {\tt DAOPHOT/FIND} in {\tt MIDAS}
\citep{Stetson:94}, applying a $3.0 \sigma$ per pixel detection threshold.
The measurements of their magnitudes were conducted via point-spread-function
(PSF) fitting that is constructed from the isolated 'PSF-stars' as it
is implemented in {\tt DAOPHOT/ALLSTAR}.
The background galaxies, unresolved blends and stars contaminated by
cosmetic CCD blemishes were eliminated from the final lists, using
their peculiar ALLSTAR characteristics,
$|SHARP|>0.3$, $\chi>1.3$ \citep{Stetson:94}.
We derived an aperture correction from the 2.5
to the 10 pixels ($0\farcs5$) aperture radius by analyzing profile of
the PSF-stars.

The $F606W$ and $F814W$ instrumental magnitudes have been transformed to
standard magnitudes in the Kron-Cousins system
adopting our empirically-derived transformation:
\begin{equation}
\begin{array}{ll}
(V-I) = 1.321 \cdot (v-i) + 1.133\\
I = i + 0.059 \cdot (V-I) + 25.972
\end{array}
\end{equation}
These transformations have been calculated using
the photometry of 60 stars with $0 < (V - I) < 5.0$, detected in
both, ACS/WFC and WFPC2 images of irregular galaxy IC~10,
and the prescriptions of \citet{Holtzman:95b} for the WFPC2 data.
The photometric uncertainties of this calibration are
dominated by the accuracy of the PSF-photometry of these stars
with a median value of $0\fm03$ in $I$-band and $0\fm04$ in $V$.

In \citet{ntik:05a} and \citet{ntik:05b} we applied these
equations to the single-star photometry of the ACS/WFC images of
NGC~55 and NGC~4244.
Based on the derived magnitude of the tip of the red giant branch
method, we estimated distances to these galaxies, which agree well
with measurements obtained with WFPC2 and other instruments.

With the extinction law of \citet{Cardelli:89}
we correct all the photometry for Galactic extinction, adopting
the foreground extinction of \citet{Schlegel:98}, and assuming that
internal extinction in all galaxies is negligible in their outer (``halo'')
parts.

\section{Results}

\subsection{CM diagrams}

\begin{figure*}[tb]
\vbox{\includegraphics{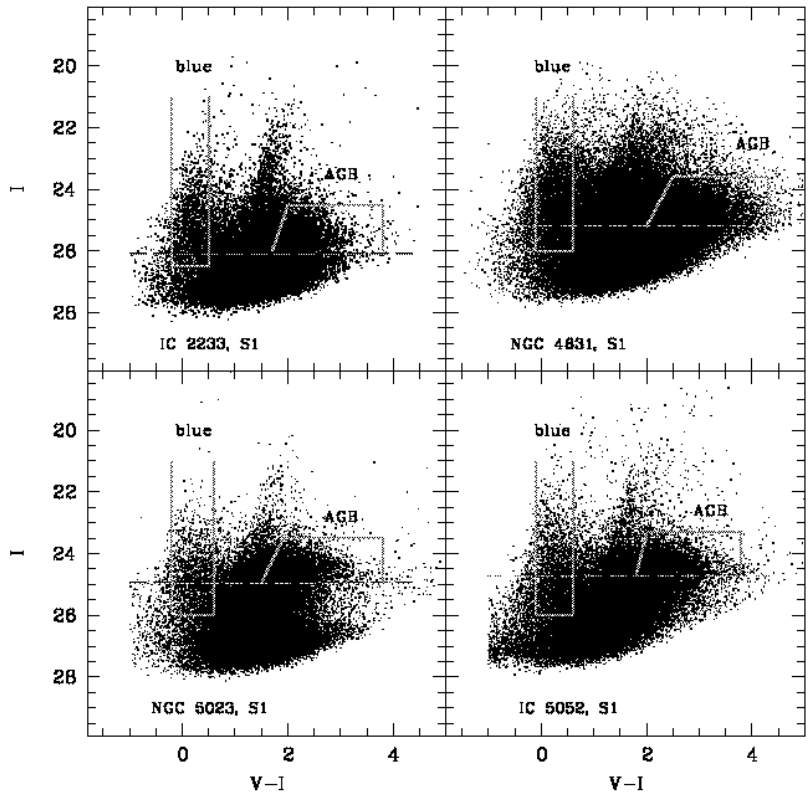}
}

\vspace{11.2cm}
\caption{The $[(V-I),I]$ color-magnitude diagram of all stars located
in the ACS/WFC fields of the galaxies. The dashed line shows the position 
of the
TRGB and the solid lines show the regions of the ``blue'', and AGB stars.
}
\label{f:CMDa}
\end{figure*}

\begin{figure*}[tb]
\vbox{\includegraphics{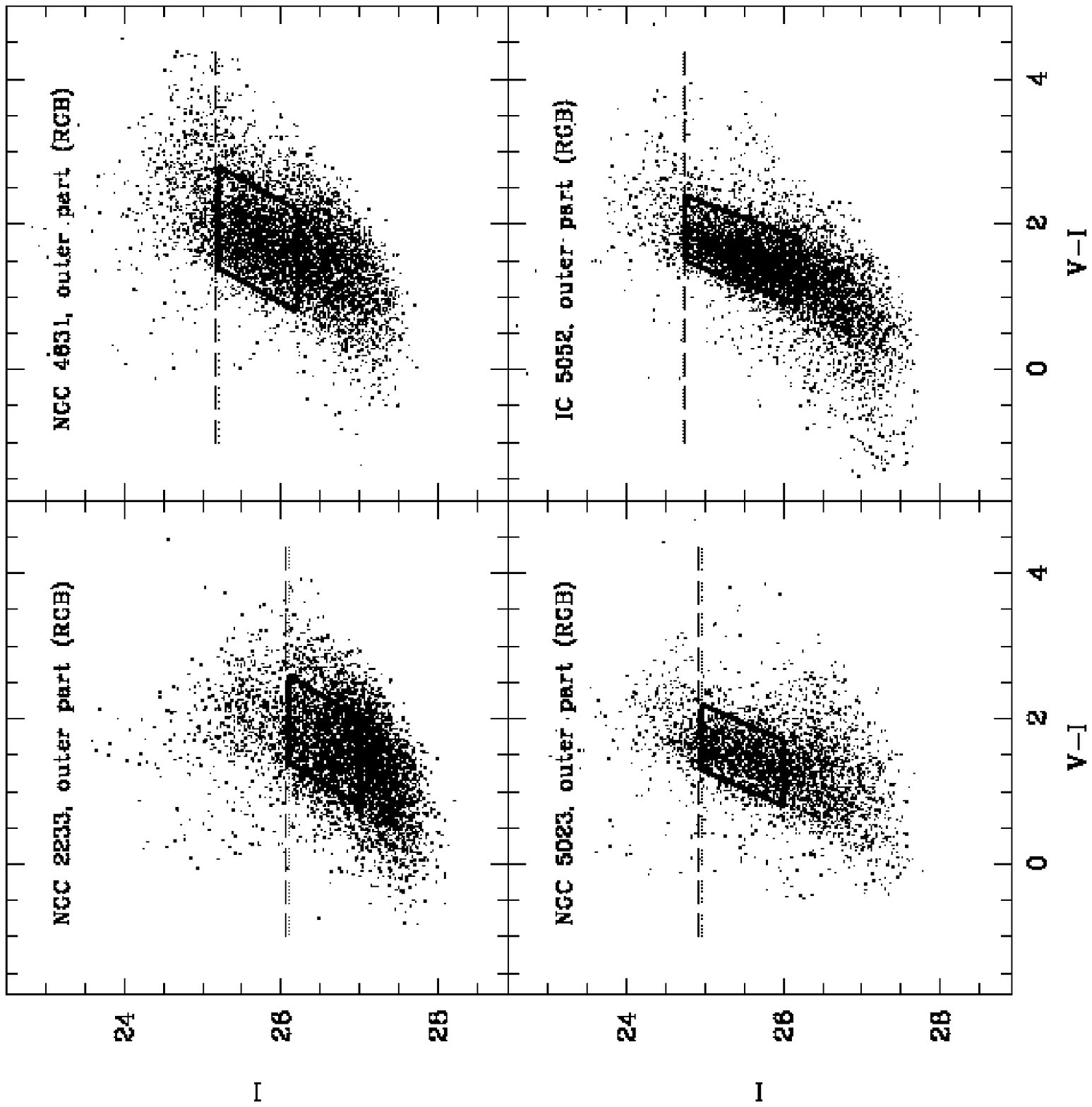}
      \includegraphics{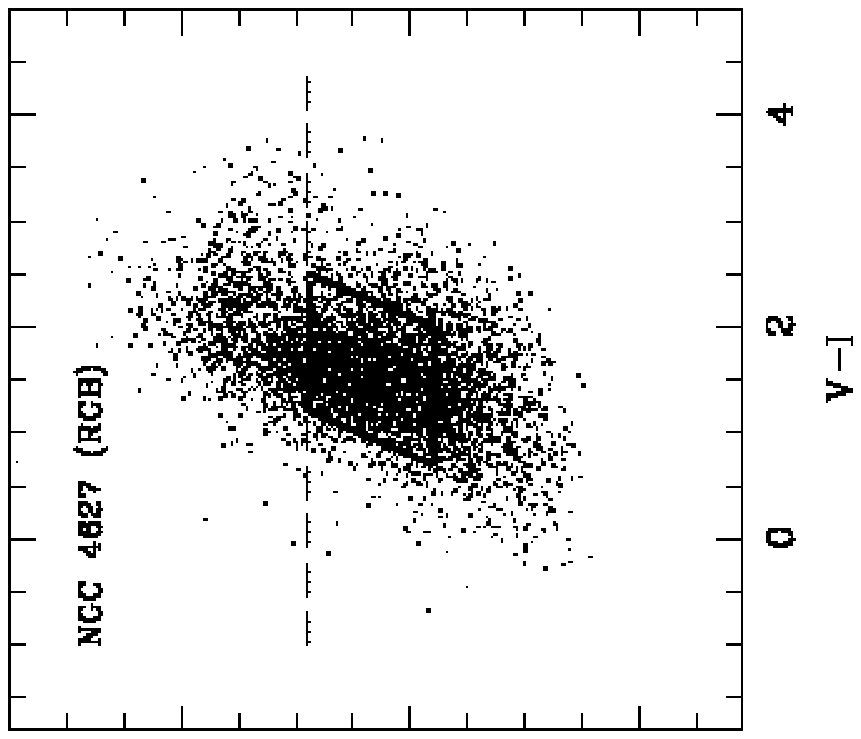}      	      
      
}

\vspace{11.2cm}
\caption{The $[(V-I),I]$ color-magnitude diagram of the stars located
above the thin disk of the galaxies. The dashed line shows the position of the
TRGB and the solid lines show the region of the RGB stars.
}
\label{f:CMDb}
\end{figure*}

The results of photometry for all stars detected in the galaxies
are presented as colour-magnitude diagrams (CMD) in
the Fig.~\ref{f:CMDa}, while
Fig.~\ref{f:CMDb} show the CMDs of the stars from the outermost 
extraplanar areas
\footnote{Note, that outer part of NGC~4631 also contain some of the stars
of the satellite dwarf Sph/Irr galaxy NGC~4627, that are shown on the
separate plot}.
The features of these diagrams resemble those of
spiral galaxies.
All the CMDs are characterized by a noticeable red
plume, while strength of the blue plume varies a lot.
The spatial variations are apparent from the varying strengths of the
blue and red plumes, and show systematic dependence from extraplanar
distance. The CMDs of the inner-disk areas show a plethora
of blue and red supergiants, main-sequence stars, and blue-loop stars.
The outermost extraplanar stellar populations are dominated by the old
stars of the red giant branch.
The dashed line indicates the position of TRGB (the tip of the red giant
branch) and above TRGB some of the stars are asymptotic giant
branch (AGB) and red supergiant (RSG) stars.

One of the ACS/WFC fields around NGC~4631, additionally includes some of
stars of the neighboring NGC~4627. Relying on their similar radial
velocities \citet{Holmberg:37} included NGC~4631/NGC~4627 in his catalog
as a galaxy pair Holmberg~442, while \citet{Arp:66} added them
in his catalog of peculiar galaxies as Arp~281.
The dwarf companion NGC~4627 is usually classified as dwarf elliptical
E4,pec (NED).
The increasing overdensity of the stars that lean towards NGC~4627
allow us to separate them
from the NGC~4631 thick disk and halo, and examine their appearance in
the CMD presented in Fig.~\ref{f:CMDb}. The RGB and AGB
stars show a well noticeable feature, that complemented by some
main-sequence stars at $-0.4<(V-I)<0.3$ and $26<I<26$, and by a few
blue supergiant candidates.
If even a small part of these blue young stars belongs to NGC~4627,
the classification of this galaxy should be altered to a
transitional dSph/Irr type. The relatively blue colour of NGC~4627
$(B-V)=0.62$ (RC3), and clear evidence of the \ion{O}{ii}-emission
around its circumnuclear area \citep{Bettoni:87} are also indicative of
the current presence of some star formation in this galaxy.

For our spatial structure analysis we used only
the stars with magnitudes above
the 50\% completeness level (i.e. the magnitude where only
half the simulated stars are identified), that occurs at
about $27.3 \leq I \leq 27.5$
above the limiting $I\sim28$~mag in these diagrams.

\subsection{Distances}

\begin{figure*}[tb]
\vbox{\includegraphics{fig06.ps}
}

\vspace{11.0cm}
\caption{Smoothed luminosity function (solid lines) and edge-detection
Sobel-filter output (dotted lines) for the red stars. The position of the
TRGB corresponds to the peak of the Sobel-filter.
}
\label{f:SLF}
\end{figure*}

The high quality of the presented data provides a valuable opportunity to
estimate distances of investigated galaxies, based on
one the most successful and reliable distance candle, the
luminosity of the TRGB \citep{Lee:93}. While being affected by
contamination of bright asymptotic giant branch (AGB) stars, statistical
(Poisson)
noise of star counts,
and dependence from the metallicity/age of RGB stars, the TRGB
method gives an estimate of distance with precision and accuracy similar
to that of the Cepheids method, $\leq10$\% \citep{Lee:93, Bellazzini:01}.

In order to estimate distances, we have used the outer extraplanar areas
in each galaxy, avoiding crowded star formation regions in a thin disk.
Since fraction of AGB-to-RGB stars steeply declines with extraplanar
distance, we can also expect of decreasing of the contamination by AGB stars
in the outermost areas. At the same time, we maintained the coverage large
enough to include a statistically significant number of RGB stars.
Following the standard technique we calculated the luminosity function
of candidate RGB stars, and applied an edge detection algorithm (Sobel
filter) to detect the location of the TRGB.
The position of the first peak in the convolved function
was used to determine the tip magnitude while the width
of the peak provides an estimate of the tip error.
The stellar luminosity functions and the results of their convolution
with Sobel filter are presented in
Fig.~\ref{f:SLF}, and reveal a sudden discontinuity which corresponds to the
TRGB. Using the method of \citet{Lee:93}, we estimate
distances to the galaxies and mean metallicity of their red
giants.
The inferred metallicity of RGB stars should be considered solely
as a local value,
due to the metallicity gradient commonly seen in spiral galaxies
\citep{Tiede:04}.
We obtain the following results for the galaxies.

\subsubsection{IC~2233}
The TRGB in IC~2233 is discerned at
$I_{\rm TRGB} = 26.09 \pm 0.08$, and the median colour of RGB stars
$(V-I)_{\rm -3.5,0}$ is $1\fm60$.
Using the calibration by \citet{Lee:93}, we obtain a metallicity
[Fe/H]$\simeq-0.93$ dex and a distance modulus $(m-M)_0=30.09$,
corresponding to $D=10.42\pm0.38$~Mpc. Note that the error in distance
includes random (residuals in the PSF fitting,
tip measurement, extinction) and systematic (RR Lyrae distance scale,
absolute magnitude of TRGB, photometric zero points) uncertainties.

\subsubsection{IC~5052}
Based on the luminosity function of RGB stars in the outer area of
IC~5052 (Fig.~\ref{f:SLF}), the TRGB is at
$I_{\rm TRGB,0}=24.74 \pm 0.07$. The median colour between
$I_0=25\fm1$ and $I_0=25\fm3$, which we use to determine
$(V-I)_{\rm -3.5,0}$, is $1\fm66$. These results correspond
to a metallicity [Fe/H]$\simeq-0.9$, and a distance $(m-M)_0=28.75$
($D=5.62\pm0.20$~Mpc).

\subsubsection{NGC~4631 \& NGC~4627}
We have performed single-star photometry for two fields in the vicinity of
NGC~4631.
One of the fields (S1) partly covers the circumnuclear area of NGC~4631
and the outer part of the satellite dwarf Sph/Irr galaxy NGC~4627, while
another area (S2) is at the larger galactocentric distance along the disk plane
(see \ref{f:Ima}). To avoid bright AGB stars of the NGC~4631 disk and
bulge, we selected stars of the north-east sector of the S1 field and south
part of the S2 field (marked by arrows).
The location of the TRGB is found by detecting a sharp edge in the
luminosity function of the red plume at
$I_{\rm TRGB,0}=25.20 \pm 0.05$ and median colour of the RGB
$(V-I)_{\rm -3.5,0}$ is $1.60$ (for stars of S1 field),
while for stars of S2 field $I_{\rm TRGB,0}$ is at $25.17 \pm 0.05$,
and $(V-I)_{\rm -3.5,0}=1.47$. These results correspond
to a metallicity [Fe/H]$\simeq-0.93$ (for S1) and
[Fe/H]$\simeq-1.31$ (for S2),
and a distance $(m-M)_0=29.29$ (S1) and $29.22$ (S2).
The distance modulus of NGC~4631, based on the average results for two
fields is $(m-M)_0=29.26\pm0.04$, that
correspond to a distance $D=7.11\pm0.13$~Mpc.
Roughly selecting stars around the neighboring NGC~4627, we
found the TRGB edge of their luminosity function at
$I_{\rm TRGB,0}=25.12$ and the RGB colour of
$(V-I)_{\rm -3.5,0}=1.38$. The derived distance of NGC~4627
$(m-M)_0=29.13$ ($D=6.70\pm0.15$~Mpc) is less certain than
the distance estimation of NGC~4631, due to the small star number counts
and unclear contamination by the NGC~4631 stars, but
it confirms their proximity to each other.

\subsubsection{NGC~5023}
The $I-band$ luminosity function of RGB stars in the outer area of
IC~5023 (Fig.~\ref{f:SLF}), has a discontinuity at
$I_{\rm TRGB,0}=24.95 \pm 0.05$ corresponding to the TRGB edge.
We used the median colour between $I_0=25\fm3$ and $I_0=25\fm6$
to determine $(V-I)_{\rm -3.5,0}$, is $1\fm35$.
These results correspond to a metallicity [Fe/H]$\simeq-1.7$,
and a distance $(m-M)_0=28.94$ ($D=6.14\pm0.15$~Mpc).

\subsubsection{Comparison to previous distance estimations}
In the context of distance verification,
it is a valuable opportunity to compare our TRGB distances
estimations with a results of other methods. We also list
distances calculated by \citet{Seth:05a} as $D_{\rm SDJ2005}$,
who used the similar TRGB method from HST data.  

There are three published distance
estimations for IC~2233:
7.7~Mpc \citep{Rossa:04}, 6.9~Mpc \citep{Wilcots:04}, and 10.6~Mpc
\citep{Tully:88}. Our distance measurement, $D=10.4$~Mpc, does agree well
with the \citet{Tully:88} determination, but significantly differs
from two other results. 

For IC~5052, we found the following previous estimations:
A distance of 6.0~Mpc reported by \citet{Seth:05a},
6.3~Mpc by \citet{ikar:04}, 7.9~Mpc by \citet{Rossa:04},
$5.9$~Mpc \citep{Ryan:03}, and $9.2$~Mpc \citep{Becker:88}. Our value is
$D=5.6$~Mpc. 

While our distance for NGC~4631, $D=7.1$~Mpc, is close to 
the \citet{Rand:93} measurement of 7.5~Mpc and 7.6~Mpc by \citet{Seth:05a}, 
it significantly disaccords with the estimation of 5.3~Mpc by \citet{Sofue:90}.

There are a few distance estimations for NGC~5023. \citet{vdKruit:82}
accepted a distance of 8~Mpc based on its redshift and 
H$_0=75$~km~s$^{-1}$~Mpc$^{-1}$. Distances of 5.6~Mpc \citep{Bottinelli:85}
and 4.8~Mpc \citep{Swaters:02} were
calculated using the Tully-Fisher relation.
A brightest star method measured by \citet{Sharina:99} gives $D=5.4$~Mpc.
Our TRGB distance is 6.14~Mpc, that is in an agreement with
$D_{\rm SDJ2005} = 6.61$ by \citet{Seth:05a}. 

\section{Stellar spatial distributions}


\begin{figure*}[tb]
\vbox{\includegraphics{fig7_11.ps}
}

\vspace{11.5cm}
\caption{The surface number density distribution (SN) of RGB
({\it filled squares}), AGB ({\it dots}) and blue
stars ({\it empty squares}) along the extraplanar $Z$-axis.
}
\label{f:NvsZ}
\end{figure*}


The regions outlined in Fig.~\ref{f:CMDa} and \ref{f:CMDb} indicate the regions
of the AGB, ``blue plum'' and RGB stars used for the analysis of their
extraplanar distribution.
Similarly to our previous studies of
the outer stellar surroundings of the disk galaxies
\citep{dio:03, ntik:05a, ntik:05b},
we apply the following method.
Using colour-magnitude diagrams, we separate different stellar populations and
analyze their spatial distribution.
By sorting stars on the different groups according to their life
expectancies, we assign probable membership of stars to the
different galaxy components.
The measurements of the stellar number density perpendicular to the galaxy
plane and search for the truncation radius (edge) of the thin and
thick disks allow us to trace farther a presence of stellar halo.
Unless otherwise stated, it is
assumed that stellar thick disk
and halo are close to axially symmetric, allowing us to extrapolate
the spatial distribution from mapping only a part of the galaxy.
Forthcoming, we will examine the case of two neighboring galaxies,
where strong tidal forces might caused an asymmetry of the stellar
structures.

The extraplanar distribution of the stars (see
Fig.~\ref{f:NvsZ}), shows a decline
toward the mid-plane of the thin disks, due to the
increase of stellar crowding and concentration of dust,
rapid growth to some extraplanar distance, followed by
steady exponential decline. The radially truncated exponential
disks are routinely observed in the disk-dominated edge-on
galaxies, relying on the surface photometry
\citep[\eg][]{Grijs:96, Grijs:01, Pohlen:04} and
stellar number counts \citep[][]{ntik:05b}.
The surface number density profiles such as ones presented in
Fig.~\ref{f:NvsZ}
show, that distribution of different stellar populations has
quite different scale-lengths and truncation radii.

The young blue stars reside mostly in the thin disk, that has a large
concentration of dust.
However, the majority of these stars are
bright enough to be detected even near the mid-plane of a thin disk,
and therefore they do not show the same drop-out as less luminous
evolved stars (AGB and RGB).

\subsection {IC~2233}

The intermediate-age AGB stars are generally more extended
than blue stars and their extraplanar distribution roughly follows the
exponential law. At extraplanar distance of $Z\simeq1.6$~Mpc their number
density drops to almost zero.
The underdensity of the AGB stars make it difficult to
obtain representative information on their outermost extend.
In our previous studies, we found that in galaxies
with larger sample of AGB stars, such as
in NGC~891, there are two
gradients of the AGB stars distribution: the steep decline of the
inner AGB components until 6~kpc, and the shallow decline of
the outer AGB component traced at least up to 11~kpc above the mid-plane
\citep{ntik:05b}.
The most numerous among detected, RGB stars have the shallowest of
all stellar number density gradients. The drop of their number
near the galactic mid-plane is due to their low luminosity.
At the extraplanar distance of $Z=1.9$~kpc the RGB stars show a break
in the exponential decline of their surface density,
that likely corresponds to a vertical (extraplanar) edge of the
thick disk.
Farther above the thick disk, a density of the RGB stars declines
with a shallower slope. By fitting the exponential law, we extrapolated
stellar halo out to $Z=4.8$~kpc. Unless there is an additional outermost
stellar
component not covered by the ACS/WFC field, the relatively small
extent of the stellar halo in comparison with apparent large axis length
(planar extent) indicates that stellar halo might be represented
by oblate ellipsoid.

\citet{Stil:02} believed that galaxies IC~2233 and NGC~2537 are a physical
pair. However, we did not find any apparent tidal deformations of the
IC~2233 thick disk, as we detected in the interacting NGC~4631.
Therefore, we assume an axial symmetry for the disk and halo components
of IC~2233.

It is valuable to compare the extraplanar size of the outermost stellar and
gaseous structures of IC~2233.
It is generally assumed that the size
of the neutral hydrogen envelopes of the disk-dominated galaxies is several
times larger than the stellar halos \citep[\eg][]{Martin:98, Swaters:02}.
Rescaled to our estimation of the IC~2233 distance, $D=10.4$~Mpc,
the extraplanar height of the \ion{H}{i} envelope is $Z(\ion{H}{i})=2.4$~kpc
according to measurements of \citet{Stil:02}, or $6.1$~kpc according to
\citet{Wilcots:04}. We detected the edge of the IC~2233 thick disk at
extraplanar distance of $3.8$~kpc, while its stellar halo is traced as
far as $9.6$~kpc. Note that the extraplanar emission-line gas
was traced out to $\sim1$~kpc from the IC~2233 mid-plane \citep{Miller:03}.

\subsection {IC~5052}

Similarly to IC~2233, the young blue stars of IC~5052 are concentrated
in the thin disk, that truncated at $\sim1$~kpc above the mid-plane.
The intermediate-age AGB stars are confined in both the thin and thick disks
and can be seen out to $Z=1.6$~kpc. The old/intermediate-age RGB stars
reveal break in their surface density at $Z=1.9$~kpc, but extend farther
with smaller surface density gradient tracing a stellar halo.
The extrapolated vertical size of the halo is $Z\simeq4.0$~kpc.

There is a reported detection
of a bright \ion{H}{i}-halo and extended emission \citep{Rossa:03},
but the exact size is unknown. In the presented \ion{H}{i} column density
contours \citep{Ryan:03}, the IC~5052 vertical structure is seen at least
until $Z=3.2$~kpc, but there are also indication on more extended hydrogen
filaments.

\subsection {NGC~4631/4627}

Since one of the ACS/WFC fields (S1) covers partly the circumnuclear
area of NGC~4631 while second one (S2) are well off-center along the
disk-plane, we can easily discern the bulge stars
from the disk and halo ones (see Fig.~\ref{f:Ima}).
The distribution of stars detected in the field S1 are symmetric, and
show a significant extent as for the young blue stars (truncation height
$Z\simeq2.0$~kpc), intermediate-age AGB stars ($Z\simeq4.0$~kpc),
and RGB stars, which spread farther than the S1 field borders.

The surface density profile of stars detected in the S2 field show quite
different distribution. Only profile of blue stars is symmetric relatively
to the mid-plane, while AGB and RGB stars show asymmetric breaks
at opposite sides of galactic plane: the north side is clearly
truncated at smaller extraplanar distance than the south one.
For the AGB stars the South-to-North sides asymmetry of this break
is $Z_{\rm South}/Z_{\rm North}(AGB)=2.8\div4.0$~kpc, and for
the RGB stars $Z_{\rm South}/Z_{\rm North}(RGB)=3.1\div>4.5$~kpc.

The peculiar distribution of the evolved AGB and RGB stars can be explained
by different possibilities:\\
{\it i)}
The interaction between NGC~4631 and NGC~4627 distorted closest (Northern)
side of the thick disk and halo, that resulted in their spread out toward
North part from galaxy plane. The spatial stretch of stellar structures
yielded to the diminishing of their surface density.
Certainly, some of the outermost stars in that area might actually belong
to NGC~4627. Another difficulty of this model is to explain the
high-degree symmetry of the distribution of blue stars relatively
to the galaxy plane, as it was found in both of the studied fields, S1 and
S2.\\
{\it ii)}
The asymmetry in the distribution of evolved stars is purely due to
the mixing of the NGC~4627 stars with NGC~4631.
This model is not supported by the data,
since the surface overdensity is detected toward the opposite
(Southern) side across the NGC~4631 plane.\\
{\it iii)}
The asymmetry in detection of the AGB stars of NGC~4631 is due to the
presence of the foreground dust clouds in NGC~4627, that significantly
dim light of the NGC~4631 stars, shifting them below the magnitude
of the TRGB.
This mechanism might yields to the observable undercounting
of AGB stars by $2.3$ times, and
requires an average interstellar extinction of about one magnitude in
$I$-band, that is possible for $\sim1$~kpc extraplanar distance
\citep{Golla:96, Martin:01}.
The RGB stars are more spread out and thus less affected by dust
extinction. Therefore their
South-to-North surface densities differ less (the ratio is $3\div2$).
This model is supported by the \ion{H}{i}-feature (spur 4), plausibly
associated with NGC~4627 and apparent on the radio-images
\citep{Rand:93, Rand:94}.

We estimated the outermost vertical extension of the NGC~4631 stellar
halo as $\approx8\div10$~kpc. Interestingly, from the spectroscopic
observations \citet{Martin:01} have detected the extraplanar emission line
gas around NGC~4631 out to $Z\simeq7$~kpc.

\subsection {NGC~5023}

As in the case of IC~2233 and IC~5052, the stellar surface density of
NGC~5023 stars are symmetric relatively to a disk plane within the
observed field. The young blue stars are prominent but confined to the
narrow plane ($Z=\pm0.8$~kpc) of the galaxy thin disk.
The extraplanar size of the disk-like structure of AGB stars is $1.3$~kpc,
while thick disk of the RGB stars show break at $Z=1.6$~kpc.
The halo are well traced above by RGB stars until $Z=3.2$~kpc.
NGC~5023 is the most flattened among analyzed edge-on galaxies
with length-to-height
ratio of $a/b = 9.3$ (NED). Interestingly, the
stellar halo of this galaxy indicate also of the largest flatness.
Does it indicate the rotation of the halo?
The corrugation of the NGC~5023 disk have been reported by \citet{Florido:91},
similarly to the disk of the Milky Way.
The \ion{H}{i}-disk of the NGC~5023 show a sharp edge at $Z=1.8$~kpc,
while some of the hydrogen filaments are seen at larger distances
\citep{Swaters:02}.

\section{Extraplanar colour gradients of the RGB stars}
\label{s:metall}

Metal abundances act as a tracer to the conjoint action of star formation
and dynamical mixing of stars and gas within a galaxy.
Therefore spatial abundance variations are among the most important parameters 
in any theory of galactic chemodynamical evolution.
The majority of the investigations concentrated predominantly on the radial
(in-plane) abundance gradients in the galactic disks, which have been
established in the Galaxy and other nearby disk galaxies based on the
spectroscopy of a variety (mostly of young) objects such as stellar clusters,
H{\sc ii} regions, planetary nebulae, and brightest stars
\citep[\eg][]{Shaver:83, Friel:93, Rolleston:00, Marquez:02, Andrievsky:02}.
Edge-on galaxies provide an opportunity to explore the abundance gradient as
a function of extraplanar height, that ensure minimal contamination
by stars belonging to the dominant galactic disk component.
However, only a few edge-on galaxies are close enough to
perform metallicity measurements from the spectrum of
evolved stellar populations due to their low luminosity
\citep{Reitzel:02}.
The idea to use the $(V-I)$ colour of the RGB stars comes from its much
greater sensitivity to metallicity than it has to age, making the
mean RGB colour a good stellar metallicity indicator, although there is
some degeneracy \citep{Lee:93}.
The method is based on the statistical approach and its accuracy depends
on the quality of the photometric data as well as on the number of
detected RGB
stars (including the source confusion--Galactic red dwarfs, etc...).
The advantage of this method is that red giants
can be situated at relatively large galactocentric distances,
where surface brightness is very low and spectral methods are
below the sensitivity limits.\footnote{
Note, that the metallicity gradients of young and evolved stars reflects
different structures, and are the subject of the different star forming and 
kinematic effects:
{\em i} These estimations deal with different (spatial?) substructures -- thick
and thin disks.
{\em ii} The metallicity of young stars reflects rather the SFR and CEL,
while the
distribution of old stars are more depends on the effectiveness of the process
of stellar mixing.
We consider here only the metallicity of the upper
(brightest) part of the RGB (i.e. intermediate-age/old stellar populations).
}

\begin{figure}[tb]
\vbox{\includegraphics{fig12.ps}
}

\vspace{14.0cm}
\caption{The distribution of the $(V-I)$ color of the RGB stars from 0.3 to
0.7 mag below the tip of the RGB for three extraplanar zones: thin and thick 
disks ({\it solid lines}) and halo ({\it dotted line}). 
The similar morphology gradient in the RGB is visible in both of 
the edge-on galaxies, NGC~5023 and IC~5052: the red RGB stars are
more concentrated to the disk plane, while outermost extraplanar stars
are systematically bluer.
}
\label{f:metall}
\end{figure}

The number density of RGB stars varies several orders of magnitude across
the extraplanar space. To include statistically significant data, we
track colour changes of RGB stars located in three major areas:
the equatorial (thin disk) zone,
thick disk and halo. We consider only two galaxies for this study, NGC~5023
and IC~5052. NGC~4631 shows the apparent distortion of its disk, and
make it difficult to separate stars belonging to different spatial
substructures.
IC~2233 is too distant to collect enough stars for the reliable statistical
study.

Extraplanar colour distribution of RGB stars are similar in
NGC~5023 and IC~5052
(see Fig.~\ref{f:metall}) and characterized by weak negative median
$(V-I)$ gradient out to the ``halo'' area.
Zero or slightly negative vertical colour gradient
have also been found in some other previously studied edge-on disk galaxies, 
such as NGC~55 and NGC~4244 \citep{ntik:05a, ntik:05b}.
There is no apparent changes in colour gradient within the thick disk
area based on our measurements. \citet{Mould:05} found the zero or
slightly positive vertical colour gradients for the RGB stars in the
thick disks of four others local edge-on galaxies, but no data was available
for the $z>2$~kpc area of these galaxies. 
Based on the similar HST data, \citet{Seth:05b} confirmed that extraplanar 
stars show little or no colour gradient.

Note, that no correction has been made for
interstellar reddening, that might mostly affect the colours of the thin disk
stars.
Additionally due to the crowding,
the sample of objects within
a few hundreds parsecs of the disk mid-plane is likely dominated by the
outer-disk stars with large galactocentric radii.
The lack of the
inner-disk RGB stars, as well as age-metallicity bias
might significantly impact the cumulative colour distribution.
Thus the lack of strong colour gradient do not ruled out the
systematic chemical changes of RGB stars perpendicular to a disk plane,
as well as star to star variations in elemental abundance ratios. Even
the outermost ``halo'' RGB stars apparently show a large range of colours,
indicating that extraplanar areas have experienced a complicated star
formation history with various self-enrichment processes.

\section {Model of the stellar structures of a disk galaxy}

\begin{figure}[tb]
\vbox{\includegraphics{fig13.ps}
}

\vspace{7.0cm}
\caption{A scaled 3-D represantation of the stellar components of a typical
spiral galaxy and results of their projection as a stellar number density
for a face-on ({\em bottom plot}) and edge-on ({\em right}) galaxy.
}
\label{f:Model}
\end{figure}

Comparing the inferred parameters for the thin \& thick disk and halo
with our and other published results, we compose a scale 3-D model
of the stellar structures of a typical disk galaxy (Fig.~\ref{f:Model}).
This model relies on the star number counts in the low-inclination,
M~81 and NGC~300, and eight high-inclination disk galaxies,
NGC~55, NGC~891, NGC~4144, NGC~4244, NGC~4363, NGC~5023, IC~2233, IC~5052
\citep[][this paper]{ntik:05a,ntik:05b}. Now we have new results of
stellar photometry of four disk galaxies: \object{NGC~1311}, \object{UGC~1281}, \object{UGC~8760} and
\object{IC~1959} that to support this model of the stellar structure.
While our primary goal in this work is to constrain the spatial geometry
of the spiral galaxies, we would like to note that massive irregular
galaxies, such as IC~10 \citep{dio:03} and
M~82 \citep{ntik:05c}, show all kinds of the stellar structures, including
thick disk and halo.
The spatial distribution of stars along and perpendicular to a disk galaxy
plane are alike. The young blue stars are mostly confined to the thin
disk of the galaxy, while elder stars are more spread out with oldest
stars are traced outermost.

\section{Discussion}

\begin{table*}
\caption{Properties of Galaxies}\label{t:galres}
\small
\begin{tabular}{rcrrrrrrrrrrr}\\ \hline \hline
\multicolumn{1}{c}{Name}&
\multicolumn{1}{c}{R.A.(2000.0)}&
\multicolumn{1}{c}{DEC(2000.0)}&
\multicolumn{1}{c}{$V_h$}&
\multicolumn{1}{c}{$a \times b$}&
\multicolumn{1}{c}{$B_t$}&
\multicolumn{1}{c}{$B_t^0$}&
\multicolumn{1}{c}{$Type$}&
\multicolumn{1}{c}{$A_v$}&
\multicolumn{1}{c}{$i$}&
\multicolumn{1}{c}{$m-M$}&
\multicolumn{1}{c}{$M_{B}$} &
\multicolumn{1}{c}{} \\
\hline \\
IC2233& $08^h13^m58\fs91$& $+45\degr44\arcmin31\farcs7$ & 565  & 5.2$\times$0.6& 13.07& 11.70& SB(s)d:sp & 0.171&90& 30.09& $-$18.39& \\
IC5052& $20^h52^m01\fs63$&$-69\degr11\arcmin35\farcs9$  & 598  &  5.9$\times$0.8  & 11.16& 10.52&SBd:sp & 0.168& 90   & 28.75& $-$18.23& \\
NGC4631&$12^h42^m08\fs01$& $+32\degr32\arcmin29\farcs4$ & 606  & 15.5$\times$2.7 & 9.75 & 8.61 &SB(s)d &  0.056&  90   & 29.26& $-$20.65&\\
NGC5023&$13^h12^m12\fs60$& $+44\degr02\arcmin28\farcs4$ & 407  & 7.3$\times$0.8 & 12.85 & 11.59 &Scd & 0.060& 90   & 28.94 & $-$17.35&\\
\hline
\end{tabular}
\begin{list}{}{}
\item
The Galactic extinction correction is by \citet{Schlegel:98}.
\item
The inclination is taken from LEDA.
\item
The value of $(m-M)$ is from this paper.
\end{list}
\end{table*}

Relying on the star number counts in the disk galaxies
IC~2233, IC~5052, NGC~4631 and NGC~5023, we examine the extraplanar
spatial distribution of young and evolved stars in their disks and halos.
These  edge-on disk galaxies
show the similar morphological properties, thin and thick disks and
halo, but relative sizes of these structures are different in each galaxy.
Combining this work with results obtained for other four
large edge-on disk galaxies
\citet{ntik:05a,ntik:05c}, we suggest that most of the massive disk
galaxies have not only a thick disk but also an extended halo, consisting
mainly of evolved stars.

The stellar number density of the halo at the truncation radius (break)
of the thick disk varies a lot, and plausibly correlate with
overall flatness of a disk galaxy. The ratio of the halo-to-thick disk
size is relatively constant, $R_{\rm halo}/R_{\rm thick disk}=2.5\pm0.5$.
It might reflect the overall rotation of the galaxy, including flatten
stellar halo.

Noteworthy the significant contrast between stellar structures of two nearest
giant spiral galaxies, the Milky Way and M~31, the only systems there
possible to measure the kinematics and chemical properties of individual
stars. While both of the ``sister'' galaxies are about the same
baryonic-plus-dark masses, their stellar disks and halo differ significantly.
The outer second disk-like stellar structure of M~31 are 3-4
times more extended
than in our Galaxy and dominate over the halo stars at least
out to a de-projected galactocentric radii of about 40~kpc
\citep[\eg][]{Worthey:05,Ibata:05}.
The significant difference in the properties of the
outer stellar populations of the nearby disk galaxies, such as the Milky Way,
M~31, and others (including our dataset) plausibly suggest that the final
morphology of a
galaxy is extremely sensitive to the details of the satellite accretion
history,
as well as peculiarities of the star formation and feedback.

Such unique populations as bulge, thin and thick disks, and halo have likely
formed by different mechanisms, especially considering the differences in
ages and metallicities. The thick disk and halo
may cast light on the most ancient structure formation, so being
extremely important for the understanding of early history of galaxy
formation and evolution. If stellar halo have been formed prior to the
thick disk \citep[see a recent review of various formation scenarios
for the stellar halo by][]{Wyse:03}, the later should show both the
characteristics age and
metallicity profile of a posthalo epoch of star formation 6-12 Gyr ago
and differ significantly from the halo \citep{Mould:05}.
The addition of results from photometry of more evolved stars
(\eg Horizontal Branch
and old Main-Sequence Turn-Off stars), as well as
from kinematic and chemistry of resolved stars are
required to refine various formation scenarios.

\section{Summary}

On the basis of the star count method, we have analyzed extraplanar stellar
structures in four large high-inclination galaxies.
The main results can be summarized as follows:
\begin{itemize}
\item
The extended stellar thick disks and halos have been detected in
IC~2233, IC~5052, NGC~4631 и NGC~5023.
\item
There are clear differences between surface density gradients of
evolved stellar populations assigned to the thick disk and halo of these
disk galaxies, which allowed us to detect the edge of the thick disk.
The extraplanar distribution of the outermost stars indicate that stellar
halo is flattened along the extraplanar direction.
\item
Having a large, high quality photometry of the RGB stars
allowed us to estimated the distances to the galaxies from
the TRGB magnitude with high confidence. We determine a distance of
$10.42\pm0.38$~Mpc for IC~2233, $5.62\pm0.20$~Mpc for IC~5052,
$7.11\pm0.13$~Mpc for NGC~4631
($6.70\pm0.15$ for its satellite NGC~4627) and $6.14\pm0.15$~Mpc for NGC~5023.
\item
In NGC~4631, the revealed asymmetry of its stellar thick disk and halo,
plausibly is an evidence of an past ore ongoing interaction with
NGC~4627.
\item
The stellar distribution of stellar populations of these four galaxies
provide more confidence for the unified model of three-dimensional structure
of large disk galaxies, and motivate follow-up spectroscopic
study to constrain the kinematic characteristics of the stellar components.
\end{itemize}

\begin{acknowledgements}
This work has been financially supported by grant RFBR 03-02-16344.
Data from the NASA/IPAC Extragalactic Database have been used.
\end{acknowledgements}

\bibliographystyle{aa}
\bibliography{edge-on}

\end{document}